\documentclass{article}

\usepackage{amsmath}
\usepackage{amsfonts}
\usepackage{algorithmicx}
\usepackage{graphicx}
\usepackage{algorithm,algpseudocode}

\newtheorem{proposition}{Proposition}
\newtheorem{theorem}{Theorem}

\begin{document}

\title{Adaptive $l_1$-regularization for short-selling control in portfolio selection}

\author{S. Corsaro\thanks{Dipartimento di Studi aziendali e quantitativi, Universit\`a di Napoli ``Parthenope'',
Via Generale Parisi, 13, I-80133 Napoli, Italy, email: Italy stefania.corsaro@uniparthenope.it}
\and V. De Simone\thanks{Dipartimento di Matematica e Fisica,
Universit\`a della Campania ``Luigi Vanvitelli'',
Viale Lincoln, 5, I-81100 Caserta, Italy, email: valentina.desimone@unicampania.it}}

\date{}
\maketitle

\begin{abstract}
We consider the $l_1$-regularized Markowitz model, where a
$l_1$-penalty term is added to the objective function of the classical mean-variance
one to stabilize the solution process, promoting
sparsity in the solution.
The $l_1$-penalty term can also be interpreted in terms of short sales,
on which several financial markets have posed restrictions.
The choice of the regularization parameter plays a key role
to obtain optimal portfolios that meet the financial requirements.
We propose an updating rule for the regularization parameter
in Bregman iteration to control both the sparsity and the number
of short positions.
We show that the modified scheme preserves the properties
of the original one.
Numerical tests are reported, which show the effectiveness of the approach.
\end{abstract}




\section{Introduction}

In the classical Markowitz mean-variance framework \cite{Markowitz},
portfolio selection aims at the construction of an
investment portfolio that exposes investor to minimum risk providing him
a fixed expected return.
This approach was proposed by Markowitz
in his aforementioned seminal paper, where he stated that
portfolio selection strategy should provide an optimal trade-off between
expected return and risk (mean-variance approach).
In a successive work \cite{Markowitz1959},
Markowitz reinforced his theory arguing that, under certain, mild conditions,
a portfolio from a mean-variance efficient frontier
will approximately maximize the investor's expected utility. \\
Markowitz model relies on information about future, since expected returns
should actually be computed discounting future flows, that are clearly not
available. A common choice is to use historical data as predictive of the
future behavior of asset returns. This practice has certain drawbacks;
indeed, a limited amount of relevant historical data is often
available. Moreover, correlation between assets returns can lead to
ill-conditioned covariance matrices.\\
It is well known that errors in estimation of expected values
affect solutions more severely than errors on variances.
For this reason, to overcome this issue
some authors focus on minimum-variance portfolios,
which do not take into account the return constraint. We
recall \cite{DeMiguel} and references therein.
Moreover, different regularization techniques have been suggested;
a review of them can be found in \cite{Carrasco}. Among these,
penalization techniques have been considered, both for the
minimum- and the mean-variance approach.
In \cite{DeMiguel} $l_1$ and squared-$l_2$ norm constraints are proposed
for the minimum-variance criterion.
In \cite{Yen} an algorithm for the optimal minimum-variance portfolio selection
with a weighted $l_1$ and squared-$l_2$ norm penalty is presented.
In \cite{siam2015} authors regularize the mean-variance objective
function with a weighted elastic net penalty.\\
In this paper, we consider the $l_1$ mean-variance regularized model
introduced in \cite{Brodie}, where a
$l_1$-penalty term is added to promote sparsity in the solution.
Since solutions establish the amount of capital to be invested in each
available security,
sparsity means that money are invested in a few securities, the active positions.
This allows investor to reduce both the number of positions to be monitored
and the transaction costs, particularly relevant for small investors,
that are not taken into account
in the theoretical Markowitz model.
Another useful interpretation of $l_1$ regularization is related to the amount
of shorting in the portfolio;
from the financial point of view negative solutions correspond to
short sales.
In many markets, among which Italy, Germany and Switzerland,
restrictions
on short sales have been established in the last years,
thus short-controlling is desired as well.
Then, the choice of the regularization parameter is crucial
in order to provide sparse solutions, with either a limited or null
number of negative components, preserving fidelity to data.\\
In this paper we propose an iterative algorithm based on a modified Bregman iteration.
Bregman iteration is a well established method for the solution
of $l_1$-regularized optimization problems.
It has been successfully applied in different fields,
as image restoration \cite{Osher1}
matrix rank minimization \cite{rank}, compressed sensing \cite{compressed}
and finance \cite{siam2015}.
Our modification to the original scheme introduces an adaptive updating
rule for the regularization parameter in the regularized model.
The algorithm selects a value capable to provide solutions
satisfying a fixed financial target, formulated in terms of
limited number of active and/or short positions.

We show that our modified scheme preserves the properties
of the original one and is able to select a good value of the
regularization parameter within a negligible computational time.
Numerical tests confirm the effectiveness of the proposed algorithm.

The paper is organized as follows. In section \ref{sec:markowitz} we
briefly recall Markowitz mean-variance model. In section \ref{sec:bregman}
we introduce Bregman iteration for portfolio selection.
Our main results are in section \ref{sec:main}, where we introduce our
algorithm, based on a modified Bregman iteration, for the $l_1-$regularized
Markowitz model. In section \ref{sec:experiments} we validate our approach
by means of several numerical experiments. Finally, in section \ref{sec:conclusions}
we give some conclusion and outline future work.

\section{Portfolio selection model}
\label{sec:markowitz}

We refer to the classical Markowitz mean-variance framework.
Given $n$ traded assets, the core of the problem is to establish the amount
of capital to be invested in each available security.\\
We assume that one unit of capital is available and define
\[
\mathbf{w} = (w_{1},w_{2},\ldots,w_{n})^T
\]
the portfolio weight vector, that is, the amount $w_i$ is invested in the $i$-th security.
Asset returns are assumed to be stationary.
If we denote with
\[ \mathbf{\mu} = (\mu_{1},\mu_{2},\ldots,\mu_{n})^T
\]
the expected asset returns, then
the expected portfolio return is their weighted sum:
\begin{equation}\label{eq:mu}
  \sum_{i=1}^n w_i \mu_{i}.
\end{equation}
We moreover denote with $\sigma_{ij}$ is the covariance between returns of securities $i$ and $j$.
The portfolio risk is measured by means of its variance, given by:
\begin{equation*}
   V = \sum_{i=1}^n \sum_{j=1}^n \sigma_{ij} w_iw_j.
\end{equation*}
Let $\rho$ be the fixed expected portfolio return and $C$
the covariance matrix of returns.
Portfolio selection is formulated as the following quadratic constrained
optimization problem:
\begin{equation}\label{eq:problemC}
\begin{array}{l}
    \min_\mathbf{w} \mathbf{w}^TC\mathbf{w} \\
    \mathrm{s.t.} \\
    \mathbf{w}^T \mathbf{\mu} = \rho \\
    \mathbf{w}^T \mathbf{1}_n = 1,\\
\end{array}
\end{equation}
where $\mathbf{1}$ is the vector
of ones of length $n$.
The first constraint fixes the expected return, according to \eqref{eq:mu}. The second one
is a budget constraint which establishes that all the available capital is invested.
The non-negativity constraint is often
added to avoid short positions.
We do not consider it here, since we aim at controlling short positions
by tuning the regularization parameter, as it is discussed in the following. \\
Let us consider a set of $m$ evenly spaced dates
\[\mathbf{t} = (t_1,t_2,\ldots,t_m) \]
at which asset returns
are estimated and build the matrix $R\in \mathbb{R}^{m\times n}$ that contains
observed historical returns of asset $i$ on its $i$-th column.
It can be shown that problem \eqref{eq:problemC} can be stated in the following form:
\begin{equation}\label{eq:problemR}
\begin{array}{l}
    \min_\mathbf{w} \frac{\mathbf{1}}{m}\|\rho \mathbf{1}_m -R\mathbf{w}\|_2^2 \\
    \mathrm{s.t.} \\
    \mathbf{w}^T \mathbf{\mu} = \rho \\
    \mathbf{w}^T \mathbf{1}_n = 1.\\
\end{array}
\end{equation}
As the asset returns are typically correlated, the matrix $R$ could have some
singular values close to zero; therefore regularization techniques,
that add to objective function some form of a priori knowledge about the
solution, must be considered.
In this paper we consider the following $l_1$-regularized problem:
\begin{equation}\label{eq:penalty}
\begin{array}{l}
    \min_\mathbf{w} \|\rho \mathbf{1}_m -R\mathbf{w}\|_2^2+\tau \|\mathbf{w}\|_1 \\
    \mathrm{s.t.} \\
    \mathbf{w}^T \mathbf{\mu} = \rho \\
    \mathbf{w}^T \mathbf{1}_n = 1,\\
\end{array}
\end{equation}
where the $1/m$ term has been incorporated into the regularization one.
From the second constraint in \eqref{eq:penalty}
it follows that the objective function
can be equivalently written as:
\[|| \rho \mathbf{1}_m - R \mathbf{w} ||_2^2 + 2\tau \sum_{i:w_i<0} |w_i| + \tau.   \]
This form points out that $l_1$ penalty is equivalent to a penalty on short positions.
In the limit of very large values
of the regularization parameter, we obtain a portfolio with only positive weights,
as observed also in \cite{JagannathanMa}.

\section{Bregman iteration for portfolio selection}
\label{sec:bregman}
Portfolio selection can be formulated as the constrained nonlinear optimization problem:
\begin{equation}\label{eq:unc}
\begin{array}{l}
    \min_\mathbf{w} E(\mathbf{w}) \\
    \mathrm{s.t.} \\
    A\mathbf{w} = \mathbf{b},
\end{array}
\end{equation}
\noindent where
\[E(\mathbf{w})= \|\rho \mathbf{1} -R\mathbf{w}\|_2^2+\tau \|\mathbf{w}\|_1\]
\noindent is strictly convex and non-smooth due to the presence of
the $l_1$ penalty term,
 \[
A = \left(
\begin{array}{c}
\mathbf{\mu}^T \\
\mathbf{1}_n
\end{array}
\right) \in\mathbb{R}^{2 \times n}
\;\; \mbox{ and } \;\;
\mathbf{b} = \left(\rho,1\right)^T  \in\mathbb{R}^2.
\]
 One way to solve (\ref{eq:unc}) is to convert it into an
 unconstrained problem,
 for example by using a penalty function$/$continuation method,
 which approximates it by a sequence:
\begin{equation}\label{eq:con}
\begin{array}{l}
    \min_\mathbf{w} E(\mathbf{w}) + \frac{\lambda_k}{2} \|
    A\mathbf{w} - \mathbf{b} \|_2^2,\quad \lambda_k\in \mathbb{R}^+.
\end{array}
\end{equation}
\noindent It is well known that, if the $k$-th subproblem (\ref{eq:con}) has solution
$w_k$ and $\{\lambda_k\}$ is an increasing sequence tending to
$\infty$ as $k \longrightarrow \infty$, any limit point of $\{w_k\}$ is a solution
of (\ref{eq:unc}) \cite{Luenberger}. Therefore, in many problems it is necessary
to choose very large values of $\lambda_k$ and it makes (\ref{eq:con}) extremely difficult
to solve numerically.
Alternatively, Bregman iteration can be used to reduce (\ref{eq:unc}) in a
short sequence of unconstrained problems by using the {\em Bregman distance}
associated with $E$  \cite{Bregman},
where, conversely, the value of $\lambda_k$ in (\ref{eq:con}) remains constant.\\
The Bregman distance \cite{Bregman} associated with a proper convex functional
$E(\mathbf{w}): \mathbb{R}^n \longrightarrow \mathbb{R} $ at point
$\mathbf{v}$ is defined as:
 \begin{equation}\label{eq:distance}
D_E^\mathbf{p}(\mathbf{w},\mathbf{v})=E(\mathbf{w})-E(\mathbf{v})- < \mathbf{p}, \mathbf{w}-\mathbf{v} >,
\end{equation}
\noindent where $\mathbf{p} \in \partial E(\mathbf{v})$
is a subgradient in the subdifferential of $E$ at point $\mathbf{v}$
and $< .\;,\;. >$ denotes the canonical inner product in $\mathbb{R}^n$.
It is not a distance in the usual sense because it is not in general symmetric
but  it does measure closeness between $\mathbf{w}$ and $\mathbf{v}$ in the sense that if $\mathbf{u}$ lies
on the line segment $(\mathbf{w},\mathbf{v})$, then the line segment $(\mathbf{w},\mathbf{u})$ has smaller Bregman
distance than $(\mathbf{w},\mathbf{v})$ does.
At each Bregman iteration $E(\mathbf{w})$ is replaced by the Bregman distance
so a subproblem in the form of (\ref{eq:con}) is solved
according to the following iterative scheme:
\begin{equation}\label{eq:bregman1}
\left \{
\begin{array}{l}
   \mathbf{w}_{k+1} = \mathrm{argmin}_\mathbf{w} D_E^\mathbf{p_k}(\mathbf{w}, \mathbf{w}_{k}) +
   \frac{\lambda}{2} \|   A\mathbf{w} - \mathbf{b} \|_2^2,\\
   \mathbf{p}_{k+1} =  \mathbf{p}_{k}-\lambda A^T(A  \mathbf{w}_{k+1} -b)
   \in \partial E( \mathbf{w}_{k+1}).\\
\end{array}
\right .
\end{equation}
The updating rule of $\mathbf{p}_{k+1}$ is chosen according to the first-order optimality
condition for $\mathbf{w}_{k+1}$ and ensures that
$ D_E^\mathbf{p_{k+1}}(\mathbf{w}, \mathbf{w}_{k+1})$ is well defined.
Under suitable hypotheses the convergence of the sequence
$\{ \mathbf{w}_k \}$ to the solution of the constrained problem (\ref{eq:unc})
is guaranteed in a finite number of steps \cite{Osher1};
furthermore, using the equivalence of Bregman iteration with the augmented Lagrangian
one \cite{GoldsteinOsher}, convergence is proved also in \cite{conlag}.
Note that the convergence results guarantee the monotonic decrease of
$\| A\mathbf{w}_k- \mathbf{b} \|_2^2$, thus for large $k$ the constraint conditions
are satisfied to an arbitrary high degree of accuracy.
This yields a natural stopping criterion  according to a discrepancy principle.\\
Since there is generally no explicit expression for the solution of the sub-minimization
problem involved in (\ref{eq:bregman1}), at each iteration the solution is computed
inexactly using an iterative solver.
So, in the last years there has been a growing interest about inexact solution of the
subproblem involved in Bregman iteration.
In recent papers it is proved that, for many applications, Bregman iterations yield
very accurate solutions even if subproblems are not solved as accurately
\cite{Osher1,compressed,GoldsteinBressonOsher}.
In \cite{errorforgetting} convergence results are obtained for piece-wise linear
convex functionals.
In \cite{BenfenatiRuggiero} the inexactness in the inner solution is controlled by a
criterion that preserves the convergence of the Bregman iteration and its features in
image restoration.

\section{Modified Bregman iteration}
\label{sec:main}

A crucial issue in the solution of \eqref{eq:penalty} is the choice of
a suitable value for the regularization parameter $\tau$, as already pointed out.
The aim is to select $\tau$ so to realize a trade-off between sparsity and short-controlling 
(requiring sufficiently large values) and fidelity to data (requiring small values).
While the literature offers
a significative number of methods for Tikhonov regularization \cite{Vogel},
$l_1$ regularization parameter selection is often based on problem-dependent criteria
and related to iterative empirical estimates, that require a high computational cost.
In \cite{Brodie} least-angle regression (LARS) algorithm proceeds by decreasing the value
of $\tau$ progressively from very large values,
exploiting the fact that the dependence of the optimal
weights on $\tau$ is piecewise linear.

In this section, we present a numerical algorithm, based on a
modified Bregman iteration with adaptive updating rule for $\tau$.
Our basic idea for defining the rule for $\tau$ comes from the
well-known properties of the $l_1$ norm and the
following proposition \cite{Brodie}:
\begin{proposition}\label{thm:prop1}
Let $\mathbf{w}_{\tau_1}$ and $\mathbf{w}_{\tau_2}$  be solution of the $l_1$-regularized
problem \eqref{eq:penalty} with $ \tau_1 $ and $ \tau_2 $ respectively.
If some of $(\mathbf{w}_{\tau_2})_i$ are negative and all the entries in
$\mathbf{w}_{\tau_1}$ are positive or zero,  we have
$ \tau_1 > \tau_2$.
\end{proposition}
We then propose an updating rule for $\tau$ that generates an increasing sequence of values.
Our aim is to modify Bregman iteration, in order
to produce solutions satisfying a fixed financial target, defined
in terms of sparsity or short-controlling or a combination of them.

Let
\[
E_k(\mathbf{\mathbf{w}}) = \|R\mathbf{w}-\rho\|_2^2+\tau_k \|\mathbf{w}\|_1,\quad k=0,1,\ldots
\]
We now prove the main result of this paper:
\begin{theorem}\label{thm:conv}

Given $(\mathbf{w}_{k+1},\mathbf{p}_{k+1})$ provided by \eqref{eq:bregman1} applied to $E_k$, it holds
\begin{equation}\label{eq:mainresult}
\tilde{\mathbf{p}}_{k+1}= \frac{\tau_{k+1}}{\tau_k}\mathbf{p}_{k+1}+ 2\Bigl(1-\frac{\tau_{k+1}}{\tau_k}\Bigr) R^T(R\mathbf{w}_{k+1}-\rho) \in\partial E_{k+1}(\mathbf{w}_{k+1}).
\end{equation}
\end{theorem}
\noindent\emph{Proof.} It holds $\mathbf{p}_{k+1} \in\partial E_{k}(\mathbf{w}_{k+1}) =
\partial (\tau_k \| \mathbf{w}_{k+1}\|_1) + \nabla \Bigl( \|R\mathbf{w}_{k+1}-\rho\|_2^2\Bigr)$,
thus a vector $\mathbf{q}_{k+1}\in \partial (\tau_k \| \mathbf{w}_{k+1}\|_1)$ exists such that
\[
 \mathbf{p}_{k+1} = \mathbf{q}_{k+1} + 2 R^T(R\mathbf{w}_{k+1}-\rho).
\]
It follows that $\mathbf{q}_{k+1} = \mathbf{p}_{k+1} - 2  R^T(R\mathbf{w}_{k+1}-\rho) \in \partial (\tau_k \| \mathbf{w}_{k+1}\|_1)$.
It is easy to verify that:
\[
\frac{\tau_{k+1}}{\tau_k}\mathbf{q}_{k+1} \in \partial (\tau_{k+1} \| \mathbf{w}_{k+1}\|_1).
\]
Then $\frac{\tau_{k+1}}{\tau_k}\mathbf{q}_{k+1}  + 2 R^T(R\mathbf{w}_{k+1}-\rho) \in \partial E_{k+1}(\mathbf{w}_{k+1})$,
which completes the proof.
\hfill{$\Box$}\\

\noindent
We propose the following modified Bregman iteration:
\begin{equation}\label{eq:mbregman1}
\left \{
\begin{array}{l}
   \mathbf{p}_k =  \frac{\tau_{k}}{\tau_{k-1}} \mathbf{p}_k + 2\Bigl(1-\frac{\tau_{k}}{\tau_{k-1}}\Bigr) R^T(R\mathbf{w}_k-\rho) ,\\[2mm]
   \mathbf{w}_{k+1} = \mathrm{argmin}_\mathbf{w} D_{E_k}^\mathbf{p_k}(\mathbf{w}, \mathbf{w}_{k}) +
   \frac{\lambda}{2} \| A\mathbf{w} - \mathbf{b} \|_2^2,\\[2mm]
   \mathbf{p}_{k+1} = \mathbf{p}_{k}-\lambda A^T(A  \mathbf{w}_{k+1} -b),\\
   \tau_{k+1} = h(\tau_{k})\\
\end{array}
\right .
\end{equation}
\noindent  where $h: \Re^+ \longrightarrow \Re^+ $
is an increasing, bounded function.\\
Note that relation \eqref{eq:mainresult} in Theorem \ref{thm:conv} guarantees that the iterative scheme
(\ref{eq:mbregman1}) is well defined, thus preserves the properties of the original one.\\
In this paper we choose a multiplicative form for the function $h$.
We set $\tau_{k+1} = \eta_{k+1} \tau_k, $
 where $ \eta_{k+1}$ depends on $\mathbf{w}_{k+1}$ according
 to the financial target, as shown in Algorithm \ref{alg:algo}.
Note that we are not ensured that a finite value of $\tau$ exists
that satisfies the financial target, thus we force $h$ to be bounded
by setting a maximum value $\tau_{max}$.
If the financial target is met at a certain step, then $\eta_k=1$ for all successive iterations.
Conversely, $\tau$ is set to $\tau_{max}$. In any case, there exists
an iteration $\bar{k}$ such that $\tau_k$ remain fixed at a value $\bar{\tau}$
for $k\geq \bar{k}$.
\begin{algorithm}
\caption{Modified Bregman Iteration for portfolio selection}
\label{alg:algo}
\begin{algorithmic}
\State Given $\tau_0>0$, $\tau_{max}$, $\lambda$, $ \theta>1$ \% Model parameters
\State Given $n_{short}$, $n_{act}$ \% Financial target parameters
\State{$k:=0$}
\State{$ \mathbf{w}_0:= \mathbf{0},  \mathbf{p}_0:= \mathbf{0}, \tau_{-1} := \tau_0, $}
\While{``stopping rule not satisfied''}
\State{ $\mathbf{p}_k =  \frac{\tau_{k}}{\tau_{k-1}} \mathbf{p}_k +
\Bigl(1-\frac{\tau_{k}}{\tau_{k-1}}\Bigr) R^T(R\mathbf{w}_k-\rho)$}
\State{ $\mathbf{w}_{k+1} = \mathrm{argmin}_\mathbf{w} D_{E_k}^\mathbf{p_k}(\mathbf{w}, \mathbf{w}_{k}) +
\frac{\lambda}{2} \| A\mathbf{w} - \mathbf{b} \|_2^2$}
\State{ $\mathbf{p}_{k+1} = \mathbf{p}_{k}-\lambda A^T(A  \mathbf{w}_{k+1} -b)$}
\State{$W^-_{k+1} = \{ i   : (\mathbf{w}_{k+1})_i< 0\}$}
\State{$W^a_{k+1} = \{ i   : (\mathbf{w}_{k+1})_i \not = 0\}$}
\If {$|W^-_{k+1}|> n_{short}$ or $|W^a_{k+1}| > n_{act}$}
\State $\eta_{k+1} =\theta$
\Else
\State $\eta_{k+1} = 1$
\EndIf
\State{ $\tau_{k+1} = \min\{\eta_{k+1}\tau_{k},\tau_{max}\}$ }
\State{ $k := k+1$}
\EndWhile
\end{algorithmic}
\end{algorithm}

\begin{theorem}\label{thm:conv2}
Let $\bar{\tau}$ be the regularization parameter value produced by the Algorithm \ref{alg:algo}
at step $\bar{k}$.
Suppose that at a certain step $ k\geq \bar{k}$
the iterate
$\mathbf{w}_k$ satisfies $A\mathbf{w}_k=b$.
Then $\mathbf{w}_k$ is a solution to the constrained problem
\begin{equation}\label{eq:result1}
\begin{array}{l}
    \min_\mathbf{w} E_{\bar{k}}(\mathbf{\mathbf{w}})\\
    \mathrm{s.t.} \\
    A\mathbf{w} = \mathbf{b}.
\end{array}
\end{equation}

\end{theorem}
\noindent\emph{Proof.}
We note that $\tau_k = \bar{\tau} \; \forall \; k\geq \bar{k}$,
thus the objective function is fixed for $k\geq \bar{k}$.
Therefore, the proof follows the proof of Theorem 2.2 in \cite{GoldsteinOsher}.\\

This result shows that if the sequence provided by
Algorithm \ref{alg:algo} converges in the sense of $\lim_{k \longrightarrow \infty} ||A\mathbf{w}_{k}-b||_2=0$,
then the iterates $\mathbf{w}_k$ will get arbitrarily close to a solution
to the original constrained problem with $\tau = \overline{\tau}$.

\section{Experimental results}
\label{sec:experiments}

In this section, we discuss some computational issues and
show the effectiveness of Algorithm \ref{alg:algo} for
solving the regularized portfolio optimization problem \eqref{eq:penalty}.\\
In Algorithm \ref{alg:algo} we set $\lambda=1,\; \tau_0= 2^{-5}, \; \tau_{max} = 1$ and
$\theta = 2$.
Iterations are stopped as soon as $\|A\mathbf{w}_k-b\|_2\leq Tol$ with $Tol=10^{-4}$
that, from the financial point of the view, guarantees constraints at a sufficient accuracy.
We implement the Fast Proximal Gradient method with backtracking stepsize rule (FISTA) \cite{Beck2}
to solve the unconstrained subproblem at each modified Bregman iteration in Algorithm \ref{alg:algo}.
FISTA is an accelerated variant of Forward Backward (FB) algorithm,
built upon the ideas of G$\ddot{u}$ler \cite{Guler} and Nesterov \cite{Nesterov}.
Note that FB is a first-order method for minimizing objective functions
$ F(x) \equiv f(x) + g(x)$,  where $g:\mathbb{R}^n\rightarrow \mathbb{R}$ is a proper,
convex, lower semicontinuous function with $\mathsf{dom}(g)$ closed,
$f:\mathbb{R}^n \rightarrow \mathbb{R}$ is convex and $\nabla f$ is $L$-Lipschitz continuous.
It generates a sequence $(x_n)_{n \in N}$ in two separate stages;
the former performs a forward (explicit) step which involves only $f$,
while the latter performs a backward (implicit) step involving a proximal
map associated to $g$ \cite{Beck}.
In our case we set
$f=\|\rho \mathbf{1} -R\mathbf{w}\|_2^2 - <\mathbf{p}, \mathbf{w}> + \frac{\lambda}{2} \| A\mathbf{w} - \mathbf{b} \|_2^2$
and $g=\tau \|\mathbf{w}\|_1$,
then the proximal map of $g$ is the simple and explicit Soft threshold operator:
\[Prox_g(w_i)=sgn(w_i) \left ( |w_i|-min \{|w_i|,\tau\} \right ). \]
Inner iterations are stopped when the relative difference in Euclidean norm between
two successive iterates is less than $Tol_{Inn}=10^{-4}$.
All our experiments, some of which are reported in the following, show that it is
not worth to require a great accuracy to the inner solver.\\
The tests have been performed in Matlab R2015a (v.~8.5, 64-bit) environment,
on a six-core Xeon processor with
24 GB of RAM and 12 MB of cache memory, running
Ubuntu/Linux 12.04.5.
We compare our optimal portfolios with the evenly weighted one
(the \emph{naive} portfolio), usually taken as benchmark in literature \cite{naive}.
This essentially for three reasons: it is easy to implement, many
investors still use such simple rule to allocate their wealth across assets
and it allows one to diversify the risk.\\
We evaluate our approach observing the out-of-sample performances of
optimal portfolios as in \cite{DeMiguel}.
This means that for each T-years period of asset
returns, we use historical series to solve \eqref{eq:penalty}; the target return $\rho$
is fixed to the average return provided by the naive portfolio in those years.
The optimal solution obtained in this way is used to build a portfolio that is retained
for one year. We continue this process by moving one year ahead until we reach the end of the
period, ending with a series of out-of-sample portfolios.
We then compare the so obtained average return $\hat{\rho}$
and standard deviation values $\hat{\sigma}$ with the
corresponding ones of the naive portfolio.
We moreover compute the \emph{Sharpe ratio} $SR=\hat{\rho}/\hat{\sigma}$:
since one would desire great return and small variance values,
the Sharpe ratio can be taken as reference value for the comparison.
We present the results on three test problems; the first and the second
one come from Fama and French database\footnote{data available at \\
\texttt{http://mba.tuck.dartmouth.edu/pages/faculty/ken.french/data$\_$library.html$\#$BookEquity}},
used in \cite{Brodie}. We obtain comparable results in terms of optimal
portfolio Sharpe ratio.
The last test problem is built on data from Italian market.

\subsection{Test 1: FF48}
\begin{figure}[htbp]
  \centering
 \begin{tabular}{c}
   \includegraphics[width=12cm]{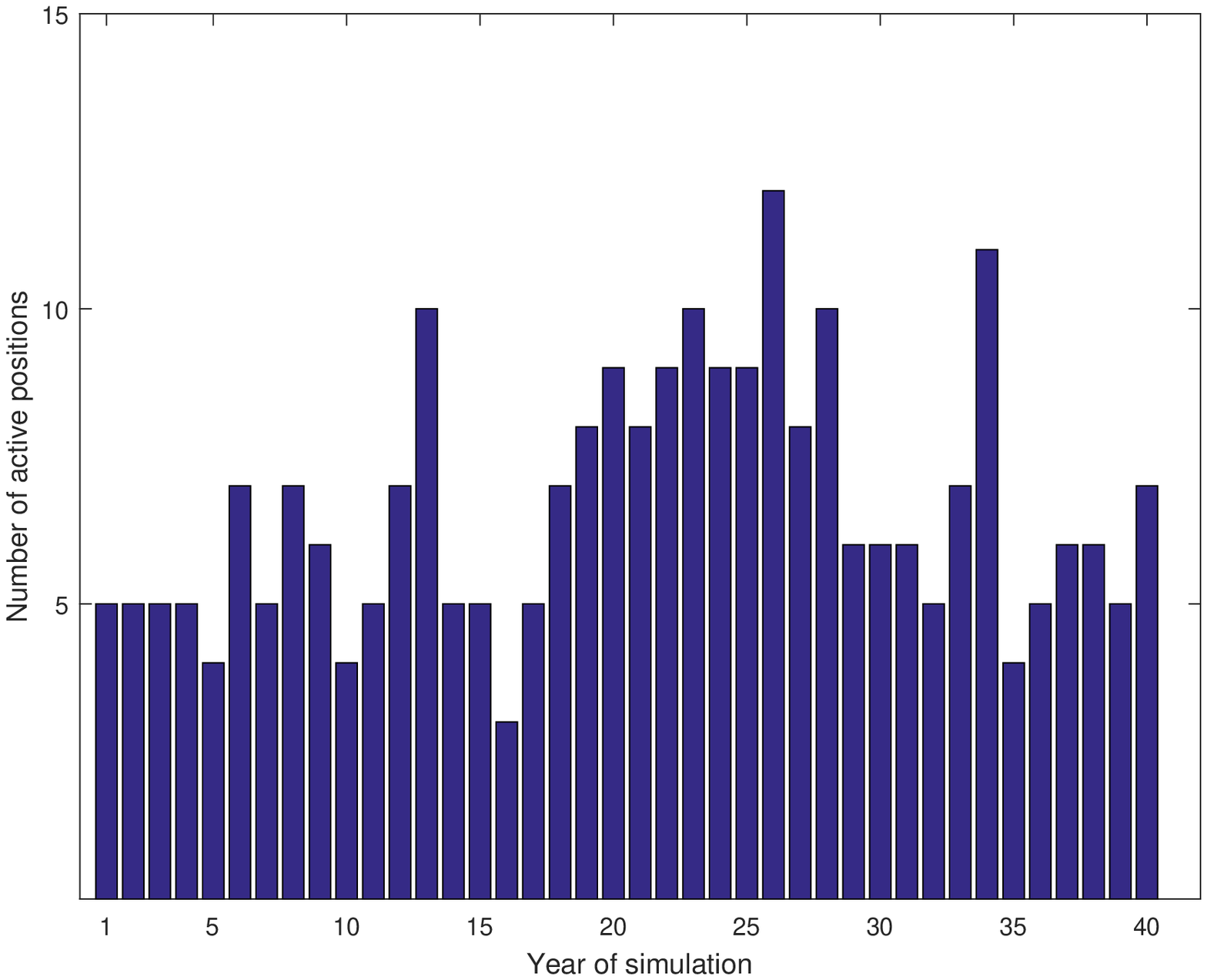} \\
   \includegraphics[width=12cm]{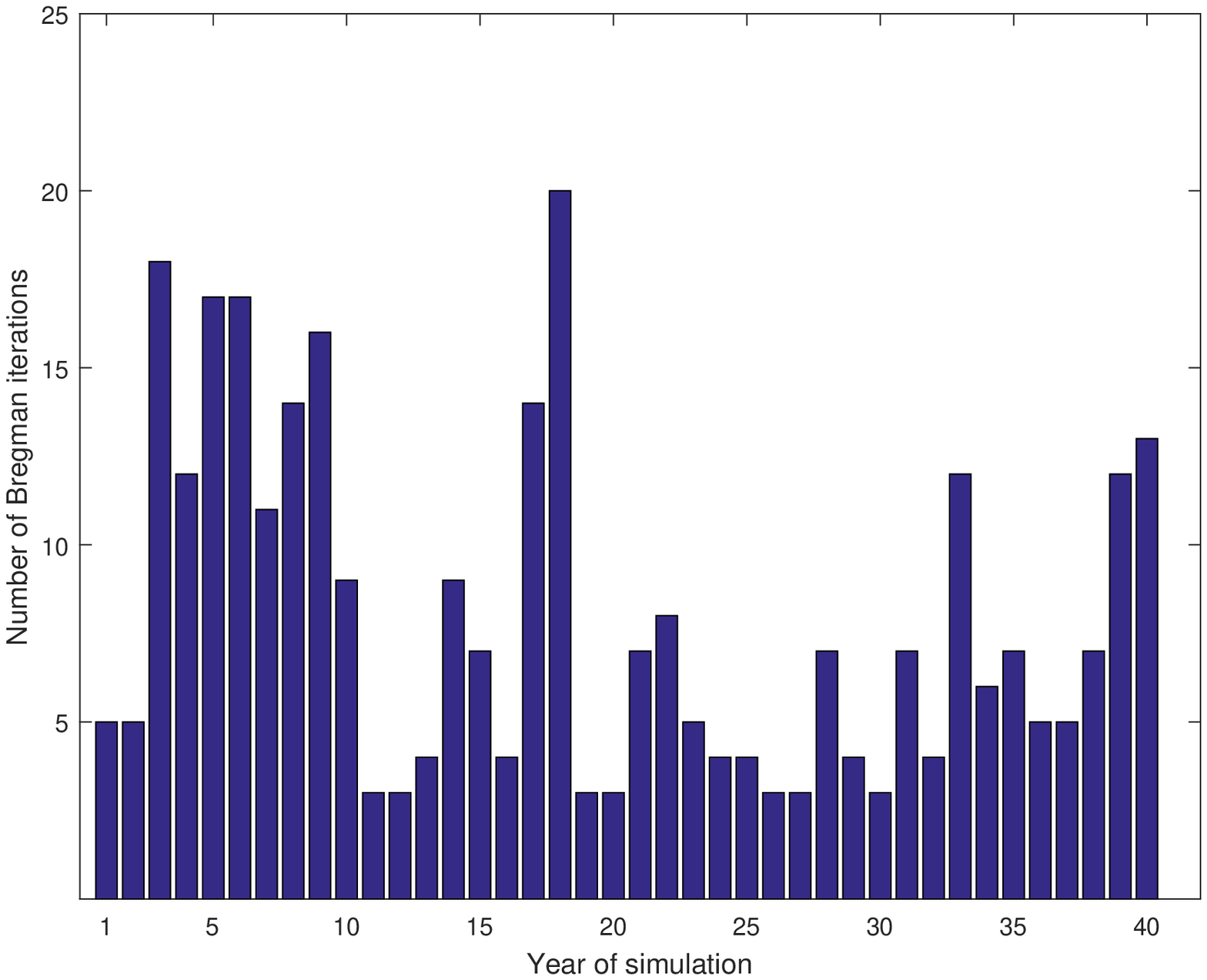} \\
 \end{tabular}
  \caption{Optimal portfolio for FF48, with $n_{short} = 0, n_{act} = 48$.
  Top: active positions.
   Bottom: number of modified Bregman iterations.}
  \label{fig:grafici48}
\end{figure}

\begin{figure}[htbp]
  \centering
 \begin{tabular}{c}
   \includegraphics[width=10cm]{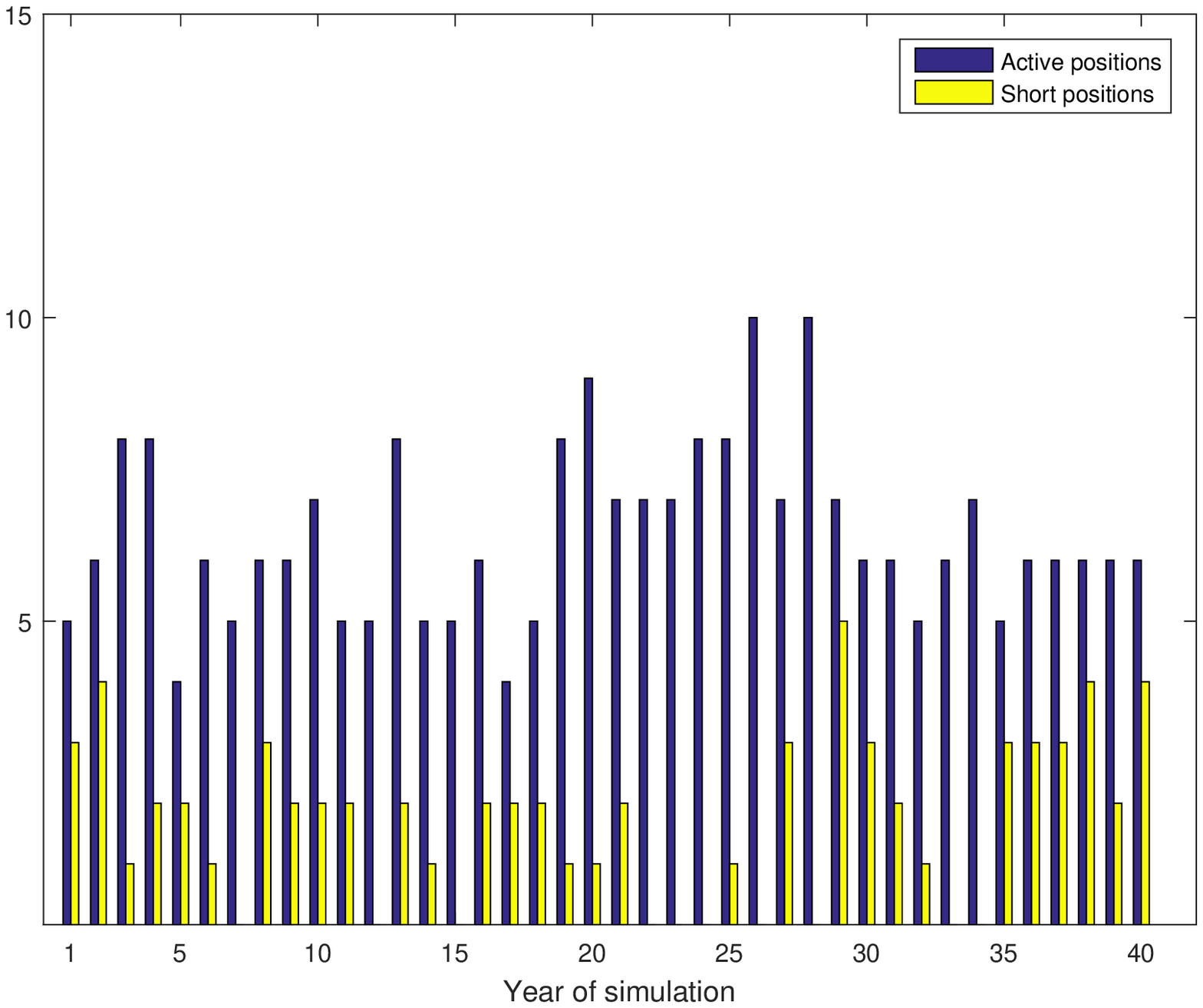} \\
   \includegraphics[width=10cm]{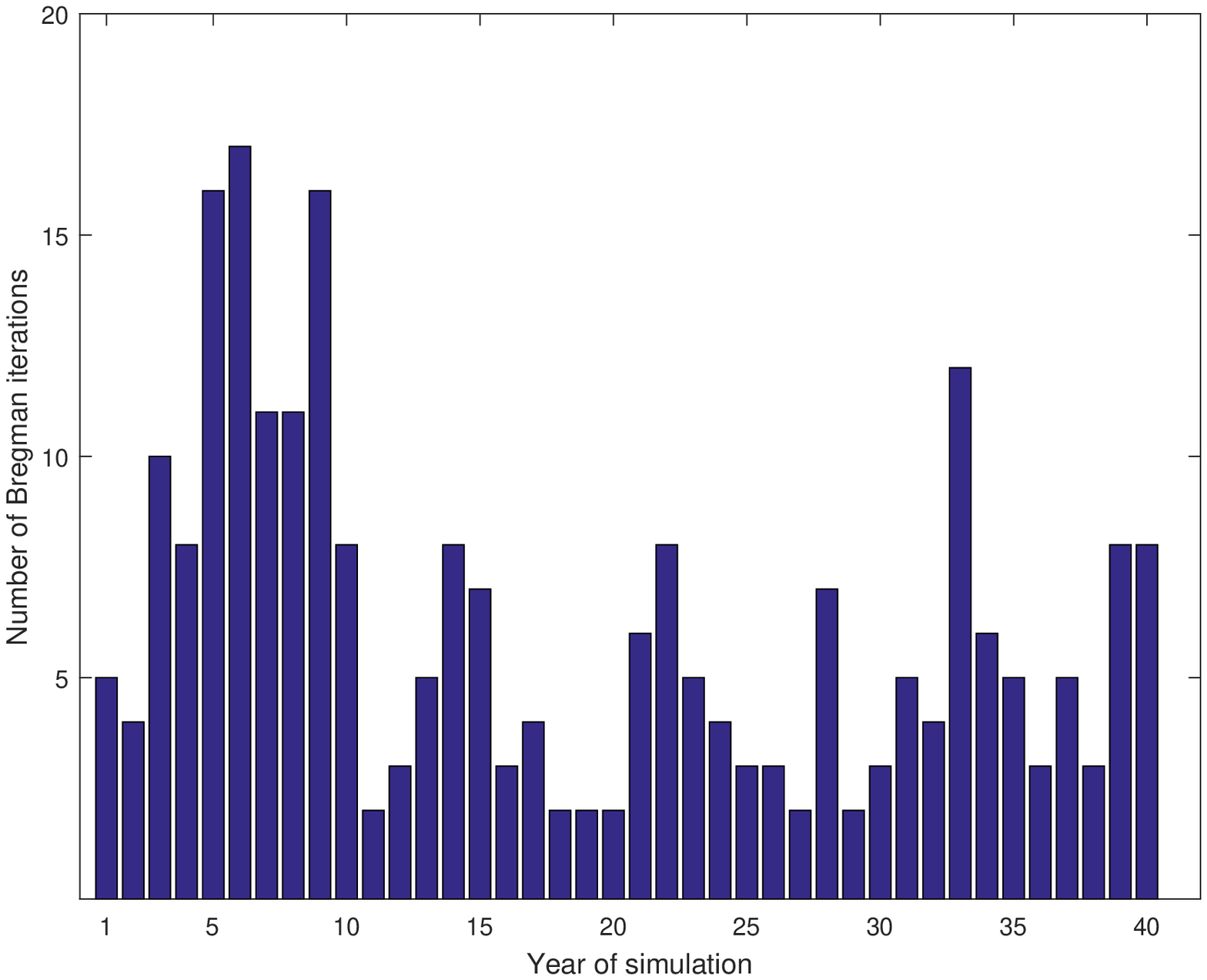} \\
 \end{tabular}
  \caption{Optimal portfolio for FF48, with $n_{short} = 48, n_{act} = 10$.
  Top: active and short positions in optimal portfolios.
   Bottom: number of modified Bregman iterations.}
  \label{fig:grafici48bis}
\end{figure}

We consider
the first database - FF48 - which contains monthly returns
of 48 industry sector portfolios from July $1926$ to December $2015$.\\
Using data from $1970$ to $2015$, we construct optimal portfolios and analyze
their out-of-sample performance. Starting from July $1970$, we use the $T=5$-years
so $40$ optimal portfolios are built, until June $2015$. Portfolios in FF48 exhibit
moderate correlation, indeed the condition number of $C$ is $O(10^4)$ for all simulations.
We tested difference values of $Tol_{Inn}$; all our experiments,
show that lower values of $Tol_{Inn}$ do not improve results, thus we show results
obtained for $Tol_{Inn}=10^{-4}$. \\
In table \ref{tab:FF48}, for both optimal and naive
portfolio, expected return, standard deviation and Sharpe ratio are
reported, all expressed on annual basis. The optimal portfolios are no-short ones,
($n_{short} = 0$, $n_{act} = 48$),
that is, the target is to obtain positive solutions.
Values refer to average values computed over
$8$ years, grouped as described in the first column of the table. The first row
contains average values computed over the all $40$-years period of simulation.
In all cases, optimal portfolio exhibits greater values of Sharpe ratio
than the naive one.\\
\begin{table}[htbp]
\centerline{
\begin{tabular}{|c|c|c|c||c|c|c|}
  \hline
  & \multicolumn{3}{|c||}{Optimal portfolio} & \multicolumn{3}{|c|}{Naive portfolio} \\
  \hline
 Period  & $\hat{\rho}$ & $\hat{\sigma}$& SR  & $\hat{\rho}$  & $\hat{\sigma}$  & SR \\
  \hline\hline
  $1975/07-2015/06$ & $14\%$ & $38\%$ & $37\%$ & $15\%$ & $60\%$ & $26\%$ \\
  \hline
  $1975/07-1983/06$ & $22\%$ & $44\%$ & $51\%$ & $29\%$ & $63\%$ & $47\%$ \\
  $1983/07-1991/06$ & $14\%$ & $39\%$ & $37\%$ & $7\%$ & $59\%$ & $12\%$ \\
  $1991/07-1999/06$ & $14\%$ & $29\%$ & $50\%$ & $15\%$ & $50\%$ & $30\%$ \\
  $1999/07-2007/06$ & $13\%$ & $34\%$ & $38\%$ & $17\%$ & $58\%$ & $29\%$ \\
  $2007/07-2015/06$ & $7\%$ & $44\%$ & $15\%$ & $10\%$ & $69\%$ & $14\%$ \\
  \hline
\end{tabular}
}
\caption{Comparison between optimal no-short ($n_{short} = 0, n_{act} = 48$)
and naive portfolio for FF48.
Reported return and standard deviation are average values over $40$ years
(first line) and over groups of $8$ years (lines $2-6$).}\label{tab:FF48}
\end{table}
In figure \ref{fig:grafici48} we report the number of active positions
in optimal no-short portfolios (top) and the number of modified Bregman iterations (bottom)
for each year of simulation.
We note the fast convergence of the Algorithm \ref{alg:algo}, with an average number
of iterations equal to 8.
The values of $\overline{\tau}$ range between $2^{-5}$ and $2^{-2}$,
promoting sparsity (the percentage of
sparsity varies from $6\%$ to $21\%$) and positivity.\\
In table \ref{tab:FF48.2}, for the same financial target,
we report results on optimal portfolios containing at most
ten active positions ($n_{short} = 48$, $n_{act} = 10$).
Values are interpreted as in table \ref{tab:FF48}.
In all cases, optimal portfolio exhibits again greater values of Sharpe ratio
than the naive one.
\begin{table}
\centerline{
\begin{tabular}{|c|c|c|c|c|c|c|}
  \hline
  & \multicolumn{3}{|c|}{\bf{Optimal portfolio}} & \multicolumn{3}{|c|}{\bf{Naive portfolio}} \\
  \hline
Period  & $\hat{\rho}$ & $\hat{\sigma}$& SR  & $\hat{\rho}$  & $\hat{\sigma}$  & SR \\
  \hline
  $1975/07-2015/06$	&$	14\%	$&$	37\%	$&$	38\%	$&$	15\%	$&$	60\%	$&$	26\%	$	\\
  \hline \hline
  $1975/07-1983/06$ &$	21\%	$&$	41\%	$&$	49\%	$&$	29\%	$&$	63\%	$&$	47\%	$	\\
  $1983/07-1991/06$	&$	15\%	$&$	38\%	$&$	40\%	$&$	7\%	    $&$	59\%	$&$	12\%	$	\\
  $1991/07-1999/06$	&$	14\%	$&$	29\%	$&$	49\%	$&$	15\%	$&$	50\%	$&$	30\%	$	\\
  $1999/07-2007/06$	&$	12\%	$&$	33\%	$&$	37\%	$&$	17\%	$&$	58\%	$&$	29\%	$	\\
  $2007/07-2015/06$ &$	8\%	    $&$	43\%	$&$	18\%	$&$	10\%	$&$	69\%	$&$	14\%	$	\\
  \hline
\end{tabular}
}
\caption{Comparison between optimal and naive portfolio for FF48.
Optimal portfolios contain at most ten active positions ($n_{short} = 48$, $n_{act} = 10$).
Reported return and standard deviation are average values over $40$ years
(first line) and over groups of $8$ years (lines $2-6$).
}\label{tab:FF48.2}
\end{table}
In figure \ref{fig:grafici48bis} we report the number of active and short positions
in optimal portfolios (top) and the number of modified Bregman iterations (bottom)
for each year of simulation.
In this case the average number of Bregman iterations is equal to 6.
The values of $\overline{\tau}$ range between $2^{-5}$ and $2^{-3}$.
Finally in table \ref{tab:FF48.confronto} we report a comparison
with results exhibited in Table 1 of paper \cite{Brodie}.
We denote with Algorithm 1 the results produced by our optimization
procedure and with LARS the results provided in \cite{Brodie}.
We refer to the same $30$-years simulation period, with the average
taken over $5$-years for each break-out period.
 We note that our procedure of regularization parameter selection produces higher values of Sharpe ratio; since the expected return is fixed by the constraint, this means that we obtain less risky portfolios.

\begin{table}
\centerline{
\begin{tabular}{|c|c|c|c|c|c|}
  \hline
  & \multicolumn{5}{|c|}{\bf{Optimal portfolio}} \\
  \hline
   & & \multicolumn{2}{|c|}{\bf{Algorithm 1}} & \multicolumn{2}{|c|}{\bf{LARS}} \\
  \hline
Period  & $\hat{\rho}$ & $\hat{\sigma}$& SR  & $\hat{\sigma}$  & SR \\
  \hline
$1976/07-2006/06$	&$	17\%	$&$	37\%	$&$	46\%	$&$	41\%	$&$	41\%	$	\\	\hline\hline
$1976/07-1981/06$ 	&$	23\%	$&$	43\%	$&$	53\%	$&$	48\%	$&$	49\%	$	\\	
$1981/07-1986/06$	&$	23\%	$&$	36\%	$&$	64\%	$&$	41\%	$&$	57\%	$	\\	
$1986/07-1991/06$	&$	9\%	    $&$	45\%	$&$	20\%	   $&$	45\%	$&$	20\%	$	\\	
$1991/07-1996/06$	&$	16\%	$&$	21\%	$&$	76\%		$&$	26\%	$&$	62\%	$	\\	
$1996/07-2001/06$ 	&$	16\%	$&$	38\%	$&$	42\%		$&$	40\%	$&$	40\%	$	\\	
$2001/07-2006/06$ 	&$	13\%	$&$	39\%	$&$	33\%	$&$	43\%	$&$	30\%	$	\\	\hline
\end{tabular}
}
\caption{Comparison between no-short optimal portfolios for FF48 produced by Algorithm
1 and LARS in \cite{Brodie}, Table 1.
Reported return and standard deviation are average values over $30$ years
(first line) and over groups of $5$ years (lines $2-7$).
}\label{tab:FF48.confronto}
\end{table}

\subsection{Test 2: FF100}
\begin{figure}[htbp]
  \centering
 \begin{tabular}{c}
   \includegraphics[width=10cm]{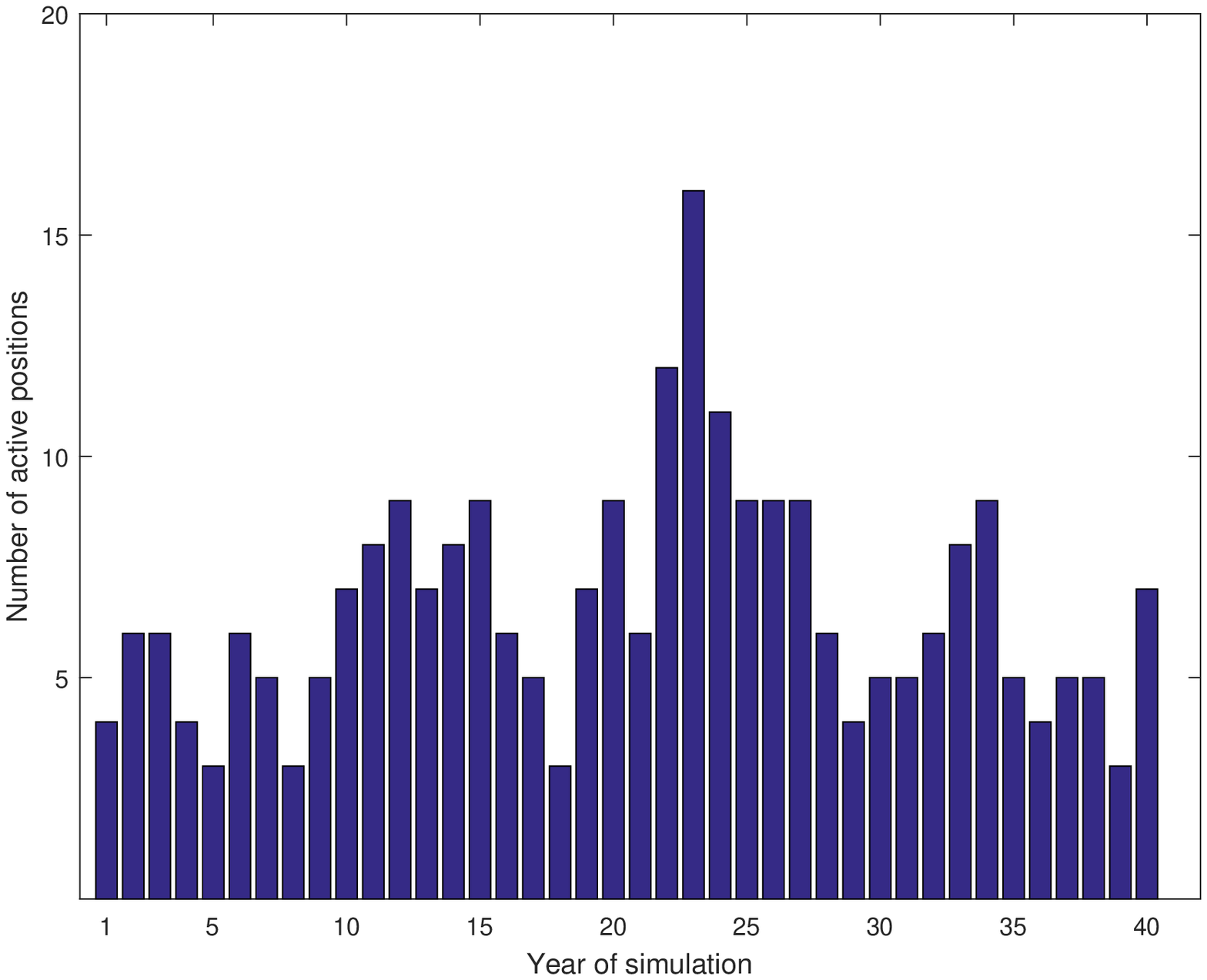} \\
   \includegraphics[width=10cm]{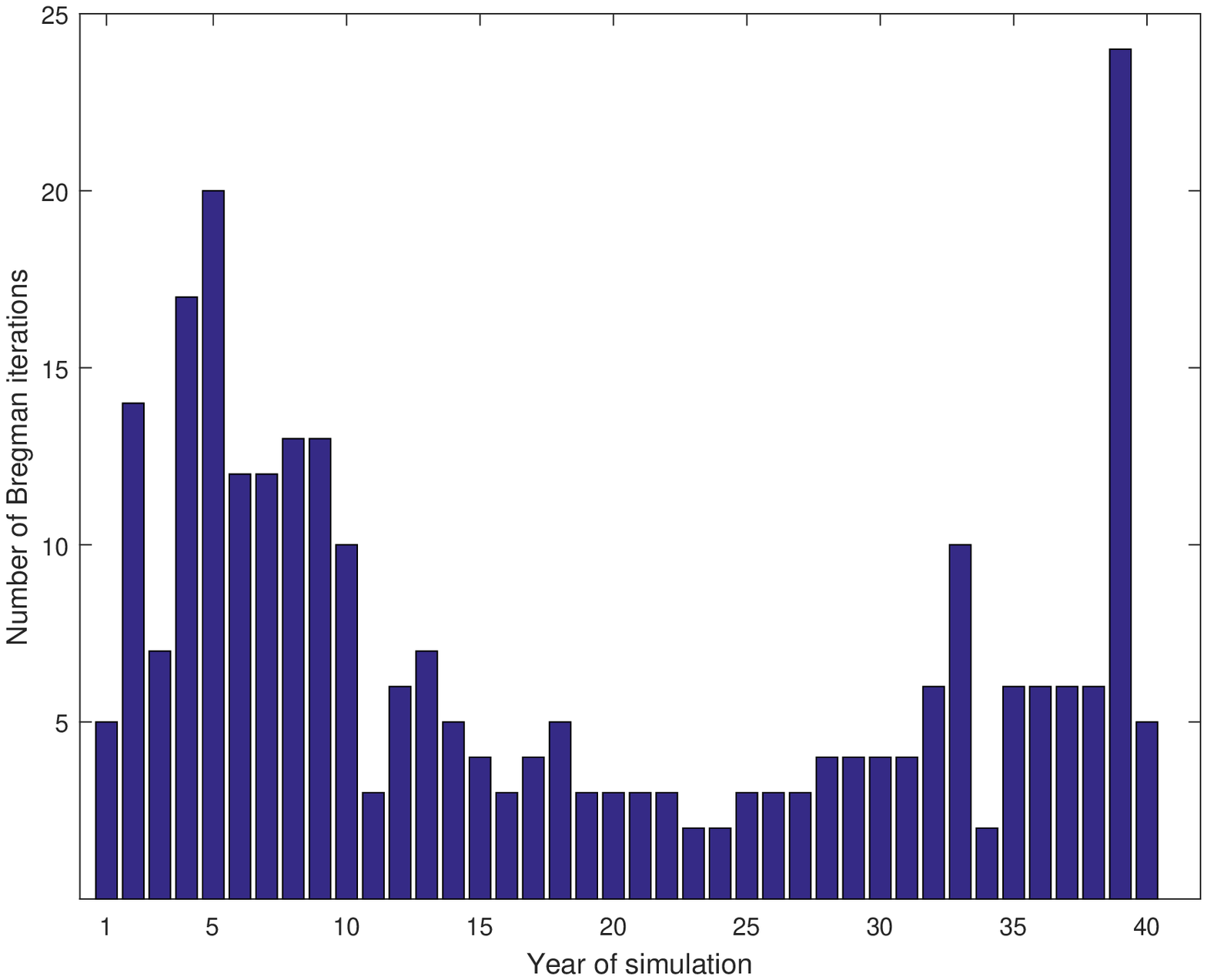} \\
 \end{tabular}
  \caption{Optimal portfolio for FF100, with $n_{short} = 0, n_{act} = 100$.
  Top: active positions.
   Bottom: number of modified Bregman iterations.}
  \label{fig:grafici100}
\end{figure}

We here show results on the second database by Fama and French
- FF100 - containing data of $100$
portfolios which are the intersections of $10$ portfolios formed on size
and 10 portfolios formed on the ratio of book equity to market equity.
Also FF100 contains monthly returns from
from July $1926$ to December $2015$.\\
We apply the same strategy as in FF48 ($T=5-$years, $40$ optimal portfolios
constructed).
Correlation values observed in FF100 are higher than in the previous
test, the conditioning of $C$ is $O(10^{18})$.\\
In table \ref{tab:FF100}, we show optimal no-short portfolios.
We report the
expected return, the standard deviation and the Sharpe ratio expressed on annual
 basis.
\begin{table}
\centerline{
\begin{tabular}{|c|c|c|c|c|c|c|}
  \hline
  & \multicolumn{3}{|c|}{Optimal portfolio} & \multicolumn{3}{|c|}{Naive portfolio} \\
  \hline
  Period  & $\hat{\rho}$ & $\hat{\sigma}$& SR  & $\hat{\rho}$  & $\hat{\sigma}$  & SR \\
  \hline
  $	07/1975-06/2015 	$ & $	14\%	$ & $	50\%	$ & $	29\%	$ & $	15\%	$ & $	57\%	$ & $	27\%	$	\\	
  \hline
  $	 07/1975-06/1983 	$ & $	18\%	$ & $	54\%	$ & $	33\%	$ & $	24\%	$ & $	58\%	$ & $	41\%	$	\\	
  $	 07/1983-06/1991 	$ & $	15\%	$ & $	54\%	$ & $	28\%	$ & $	12\%	$ & $	60\%	$ & $	20\%	$	\\	
  $	 07/1991-06/1999 	$ & $	19\%	$ & $	38\%	$ & $	49\%	$ & $	18\%	$ & $	45\%	$ & $	39\%	$	\\	
  $	 07/1999-06/2007 	$ & $	14\%	$ & $	47\%	$ & $	29\%	$ & $	14\%	$ & $	56\%	$ & $	25\%	$	\\	
  $	 07/2007-06/2015	$ & $	7\%	    $ & $	56\%	$ & $	13\%	$ & $	10\%	$ & $	67\%	$ & $	14\%	$	\\
  \hline
\end{tabular}
}
\caption{Comparison between no-short ($n_{short} = 0, n_{act} = 100$)
optimal and naive portfolio for FF100.
Reported return and standard deviation are average values over $40$ years
(first line) and over groups of $8$ years (lines $2-6$).}\label{tab:FF100}
\end{table}
On the overall period, optimal portfolio outperforms the naive one.
The values of $\overline{\tau}$ range between $2^{-4}$ and $2^{-2}$, the percentage of
sparsity varies from $4\%$ to $17\%$ (Fig. \ref{fig:grafici100}).
Note that, looking at details on each year of simulation, we observe
negative returns for both optimal and naive portfolio. For instance,
in the $8^{th}$ year of simulation, optimal portfolio produces a loss of $4\%$, the
naive one of $12\%$. In the $10^{th}$ year the losses are of $1\%$ and
$10\%$ respectively. This happens because almost all components in
portfolios show decreased returns.
Finally, we note that in the period $07/1975-06/1983$ naive portfolio
outperforms the optimal one. For instance, in the $3^{rd}$ year of simulation
the optimal portfolio, which contains $5$ assets ($56, 90, 91, 93, 95$),
produces a gain of $8\%$, versus a gain of $18\%$
of the naive one.
This behavior is essentially due to
a drastic change in asset returns with respect to historical data.
This situation could be controlled by a dynamic asset allocation strategies, for which
at the beginning of each period during
the investment horizon, the investor can freely rearrange the portfolio,  but it isn't the aim of this paper.
We finally report also in this case a comparison with results exhibited in \cite{Brodie},
Table 3. We note that we obtain
higher values of the Sharpe ratio.

\begin{table}
\centerline{
\begin{tabular}{|c|c|c|c|c|c|}
  \hline
  & \multicolumn{5}{|c|}{\bf{Optimal portfolio}} \\
  \hline
   & & \multicolumn{2}{|c|}{\bf{Algorithm 1}} & \multicolumn{2}{|c|}{\bf{LARS}} \\
  \hline
Period  & $\hat{\rho}$ & $\hat{\sigma}$& SR  &  $\hat{\sigma}$  & SR \\
  \hline
$1976/07-2006/06$	&$	16\%	$&$	48\%	$&$	33\%	$&$	53\%	$&$	30\%	$	\\	\hline\hline
$1976/07-1981/06$ 	&$	12\%	$&$	54\%	$&$	22\%		$&$	59\%	$&$	21\%	$	\\	
$1981/07-1986/06$	&$	24\%	$&$	44\%	$&$	55\%		$&$	49\%	$&$	49\%	$	\\	
$1986/07-1991/06$	&$	10\%	$&$	61\%	$&$	16\%	$&$	65\%	$&$	15\%	$	\\	
$1991/07-1996/06$	&$	19\%	$&$	29\%	$&$	66\%		$&$	31\%	$&$	61\%	$	\\	
$1996/07-2001/06$ 	&$	18\%	$&$	52\%	$&$	35\%	$&$	52\%	$&$	35\%	$	\\	
$2001/07-2006/06$ 	&$	11\%	$&$	49\%	$&$	22\%		$&$	55\%	$&$	21\%	$	\\	\hline
\end{tabular}
}
\caption{Comparison between no-short optimal portfolios for FF100 produced by Algorithm
1 and LARS in \cite{Brodie}, Table 3.
Reported return and standard deviation are average values over $30$ years
(first line) and over groups of $5$ years (lines $2-7$).}\label{tab:FF100.confronto}
\end{table}

\subsection{Test 3: IT72}
\begin{figure}[htbp]
  \centering
 \includegraphics[width=10cm,angle=-90]{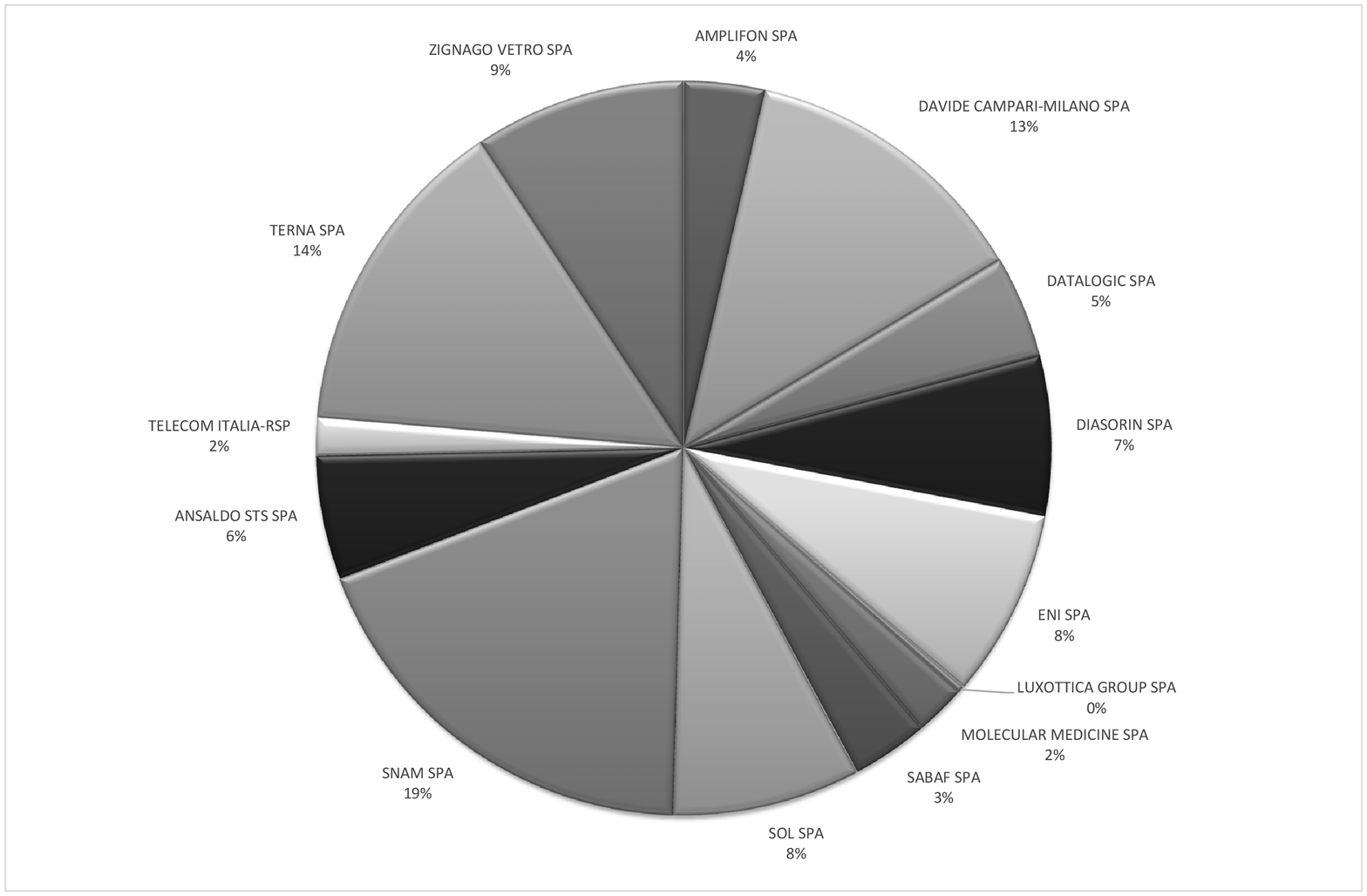}
  \caption{Optimal portfolio on Italian market equities. Built on
  monthly historical returns of $72$ equities from September $2009$ to August $2016$.}
  \label{fig:torta}
\end{figure}

We here consider a portfolio constructed on
real data from Italian market. It considers the monthly returns of $72$ equities,
from September $2009$ to August $2016$. Assets are reported in table \ref{tab:cognato};
$25$ assets are included in the FTSE MIB index computation.
The FTSE MIB is the primary benchmark Index for the Italian equity markets.
The Index is comprised of highly liquid, leading companies across different
sectors, indeed
it captures about the 80\% of the domestic market capitalization.
The FTSE MIB is computed on $40$ Italian equities and seeks to replicate the
broad sector weights of the Italian stock market. \\
\begin{table}
\tiny{
\begin{tabular}{|l|l|l|}
\hline
A2A SPA	&	EI TOWERS SPA	&	PRIMA INDUSTRIE SPA	\\
ACEA SPA	&	EL.EN. SPA	&	PRYSMIAN SPA	\\
AUTOGRILL SPA	&	ENEL SPA	&	RECORDATI SPA	\\
AMPLIFON SPA	&	ENI SPA	&	REPLY SPA	\\
ATLANTIA SPA	&	ERG SPA	&	SABAF SPA	\\
AZIMUT HOLDING SPA	&	EXOR SPA	&	SALINI IMPREGILO SPA	\\
BASICNET SPA	&	ASSICURAZIONI GENERALI	&	SAFILO GROUP SPA	\\
BIALETTI INDUSTRIE SPA	&	HERA SPA	&	SAES GETTERS SPA	\\
BANCA MEDIOLANUM SPA	&	INDUSTRIA MACCHINE AUTOMATIC	&	SIAS SPA	\\
BANCA MONTE DEI PASCHI SIENA	&	INTESA SANPAOLO	&	SOGEFI	\\
BANCO POPOLARE SC	&	INTESA SANPAOLO-RSP	&	SOL SPA	\\
BANCA POPOL EMILIA ROMAGNA	&	ITALCEMENTI SPA	&	SAIPEM SPA	\\
BREMBO SPA	&	ITALMOBILIARE SPA	&	SNAM SPA	\\
BUZZI UNICEM SPA	&	ITALMOBILIARE SPA-RSP	&	SARAS SPA	\\
BUZZI UNICEM SPA-RSP	&	LEONARDO-FINMECCANICA SPA	&	ANSALDO STS SPA	\\
CAIRO COMMUNICATIONS SPA	&	LUXOTTICA GROUP SPA	&	TELECOM ITALIA SPA	\\
CEMENTIR HOLDING SPA	&	MARR SPA	&	TELECOM ITALIA-RSP	\\
DAVIDE CAMPARI-MILANO SPA	&	MEDIOBANCA SPA	&	TOD'S SPA	\\
CREDITO VALTELLINESE SCARL	&	MOLECULAR MEDICINE SPA	&	TERNA SPA	\\
DATALOGIC SPA	&	MEDIASET SPA	&	UBI BANCA SPA	\\
DANIELI \& CO	&	MAIRE TECNIMONT SPA	&	UNICREDIT SPA	\\
DIASORIN SPA	&	PANARIAGROUP INDUSTRIE CERAM	&	UNIPOL GRUPPO FINANZIARIO SP	\\
D'AMICO INTERNATIONAL SHIPPI	&	PARMALAT SPA	&	UNIPOLSAI SPA	\\
DE'LONGHI SPA	&	BANCA POPOLARE DI MILANO	&	ZIGNAGO VETRO SPA	\\

\hline
\end{tabular}}
\caption{IT72 assets.}\label{tab:cognato}
\end{table}
Starting from September $2009$, we use the $T=6$-years data
to build the optimal portfolio from September $2015$ until August $2016$.
The conditioning of $R^TR$ is $O(10^9)$.
In figure \ref{fig:torta} we graphically show
the composition of the optimal portfolio we constructed. The optimization strategy
allocates the investor wealth on $14$ equities, with weights represented as percentage
in the figure, among which $5$ belong to the FTSE MIB set.
The result is obtained in $10$ Bregman iterations,
with $\overline{\tau} =2^{-3}$.
We note that the optimal portfolio has return and standard deviation,
on annual basis, given by $11\%$ and $34\%$ respectively. The same values for the naive
portfolio are $-14\%$ and $60\%$, thus the latter provides a loss to the investor.    \\

\section{Conclusions}
\label{sec:conclusions}

We have proposed an algorithm, which exploits the Bregman iteration method,
for the portfolio selection problem formulated
as an $l1-$regularized mean-variance model.
The choice of the regularization parameter is the key point in order to provide
solutions with either a limited or null number of negative components and/or
a limited number of active positions.
Our main contribution is the modification of the Bregman iteration,
which adaptively sets the value of the regularization parameter depending on the financial target.
It is observed that both sparsity and short-controlling are obtained for sufficiently
large values of the regularization parameter. The basic idea is then to generate an
increasing sequence of values and fix it when requirements are met.
We show that our modification to the Bregman iteration preserves the convergence of the original scheme.
Numerical experiments confirm the effectiveness of the proposed algorithm.\\
We saw in our experiments that sometimes the effectiveness of the optimization strategy
can be affected by changes in market conditions. Future work could consider
dynamic asset allocation, which involves frequent portfolio adjustments.

\section*{Acknowledgments}
This work was partially supported by FFABR grant, annuity 2017, and INdAM-GNCS project.

\end{document}